# An Architecture for Web 3.0 and the Emergence of Spontaneous Time Order

[1,3,]*Hengjin Cai, [2,3,] *Tianqi Cai

[1]School of Computer Science, Wuhan University, China

[2]Tsinghua Shenzhen International Graduate School, Tsinghua University, China

[3]Wuhan Longjin Science and Technology Inc., China

*Correspondences: hjcai@whu.edu.cn; cai.tianqi@sz.tsinghua.edu.cn

**Abstract.** In this study, we proposed an architecture for Web 3.0, which is based on the hashed interactions among user nodes that can transform bilateral trusts into collective time order, which is the major achievement of blockchain technology, without the expensive Proof of Work or the questionable Proof of Stake.

## 1 Introduction

In 1989, Tim Berners Lee proposed the World Wide Web that has completely changed the shape of the world economy. However, privacy and control of personal data has become a leading problem. Blockchain technology has emerged as basis of value exchanges networks unprecedentedly. The time order in blockchain networks is the most important property, which is the foundation of traceability, verifiability and value delivery. However, it suffers from the technical challenges of efficiency, security, energy consumption and usability. In 2014, Gavin Wood, the co-founder of Ethereum [1], proposed a new Internet operation mode of Web 3.0 with an emphasis on individualization and privacy protection [2]. That is, the information will be released and kept by the users themselves, their identities cannot be traced or disclosed, and their operations can be directly delivered instead of needing help from an intermediary organization. Since then, there have been several attempts at Web 3.0, but low efficiency and costly distributed storage remain problems.

We propose a Web 3.0 solution based on hashed interactions that emphasizes traceability and allows for free interactions and cooperation with other chains on the premise of preserving personal privacy.

## 2 Hashed Interactions Leading to Intrinsic Time Order

As the first successful application of blockchain technology, Bitcoin proposed the unspent transaction output bookkeeping method. Through the network-wide consensus of the Proof of Work, Bitcoin has formed a time order in the digital world, which is an innovation from 0 to 1. [3] [4]

The token-based bookkeeping method[5] that can support a simplified and effective Web 3.0 architecture to deal with many problems in the digital world. The proposed architecture is based on hash interactions and is characterized as follows:

i. Each chain or individual system is privately owned to the user.
ii. A user independently decides to interact with other chains, including the sending and accepting associations.

X sends an association to Y, indicating that X shares the hash value of its block summary to Y. If Y packages the hash value into its next block, Y accepts the association with X.

This approach can solve the problems of data ownership, data classification, and privacy. Furthermore, as the interactions accumulate among the nodes, the peer-to-peer network will emerge in an order without relying on network-wide consensus to order the data ontology. The resulting benefits are the forgettability of the network and the savings in energy consumption.

### 2.1 State with Established Rules

Each chain or system is user owned, indicating that all personal data are stored in privately owned devices or services, such as personal mobile devices, PCs, private clouds, among others. If users want some of their data to be visible to the network, they need to interact with other chains in a timely manner.

Figure 1 depicts a state with established rules wherein each chain in a network shows a fixed interaction pattern. For example, if each private chain interacts with at least one chain every hour, an intrinsic time sequence can be generated by a direct or indirect relation between each chain and each block. For example, the time order in Figure 1 can be described as

$$TimeOrder = \{t_{C_1}, t_{A_1}, t_{B_1}, t_{C_2}, t_{A_2}, t_{B_2}, t_{C_3}, t_{A_3}, t_{B_3}, t_{C_4}, t_{A_4}, t_{B_4}, t_{C_5}, t_{A_5}\}$$

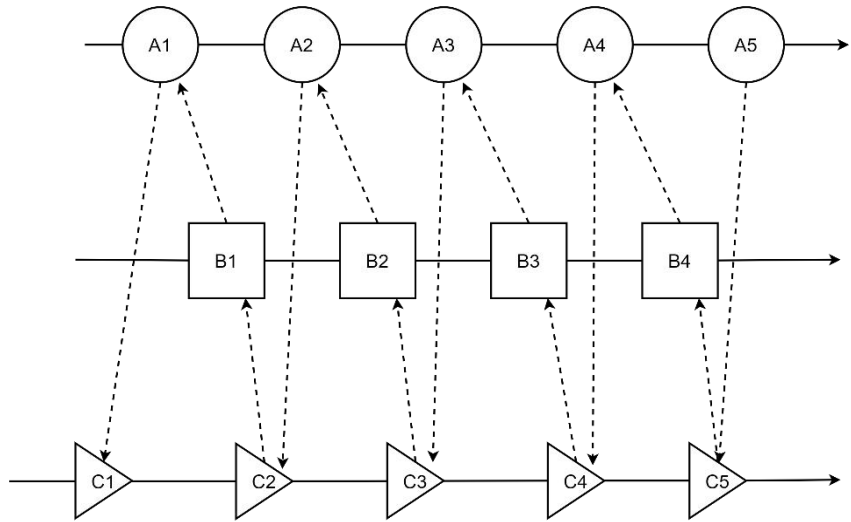

**Figure1.** Example of interactions with established rules.

## 2.2 State of Autonomous Interaction

There is no need to constrain the interaction frequency of each user. Users are free to control their own data and interact with others entirely on their own; the mechanism still works but with different time granularities or time intervals for different private chains. The various time granularities depend on how often each chain interacts with others, corresponding to the time interval between the previous block and the latter. Naturally, a chain that interacts frequently, especially with well-known chains, is more granular in time, easier to verify, and more reliable.

In multi-chain systems with different time granularities, sorting is sometimes easy, and sometimes confirming the order is difficult. Some chains may have low interaction frequency, showing large time granularity, and therefore can only be roughly sorted, which is also an acceptable result in the real world. If there are enough chains and interactions, the data can be traced, validated, and ordered.

That is, different chains have their own time granularity, which is controlled and responsible by the users themselves. If a user wants their own data to be more accurately ordered, they will interact more often with other chains, and if the user chooses to interact infrequently, then they will be responsible for the inaccuracy of their own data ordering accordingly. For example, in a state of completely autonomous interaction depicted in Figure 2, the time order has the following two possibilities.

$$TimeOrder = \{t_{A_1}, t_{C_1}, t_{B_1}, t_{A_2}, t_{B_2}, t_{C_2}, t_{A_3}, t_{C_3}, t_{A_4}, t_{B_3}, t_{A_5}\}$$

$$\text{OR } \{t_{A_1}, t_{C_1}, t_{B_1}, t_{A_2}, t_{B_2}, t_{C_2}, t_{A_3}, t_{A_4}, t_{C_3}, t_{B_3}, t_{A_5}\}$$

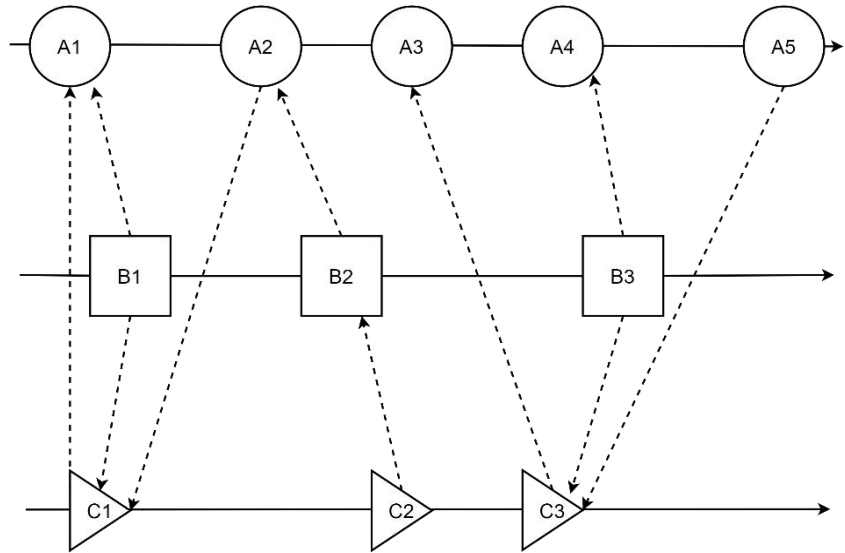

**Figure 2.** State of completely autonomous interaction.

# 3 Free Cooperation with Individuals

The order is an instinct emergence from the interactions among the nodes in the system. If we attempt to reach a fixed consensus of the order of the data ontology within the entire network at the time of the data being stored or chained, which is the circumstance in current P2P networks such as Bitcoin and Ethereum, it will require huge calculations, denoting considerable consumption of time and resources. Moreover, the initial data may be incorrect, and there will be frequent changes or patches later. Implementing a digital world with personality and efficiency is difficult if there is not enough room for understanding and reaching consensus.

The consensus is therefore separated from the data ontology in our architecture so that different requirements have matching time granularities in their respective ecology, and the relative nodes can order blocks, reach consensus, and validate data in accordance with their interactions. Each user only needs to maintain its own chain, and the data are completely private. The advantage is that the owner has a choice. Whether the data are encrypted, shared, and accepted by others are all decided by the user. The data can be classified naturally and encrypted from the ontology layer.

An interaction between different chains is actually a type of voting, which may lead to a mainstream chain at one time period and bandwagoning others at another time period. This type of governance is also closer to human thinking. People need to organize, understand, and make decisions on data, rather than make up the mind at first sight.

Accordingly, the storage and computing efficiencies will be greatly improved. As each node is responsible for its own related parts, the cost of technological renovation will be low.

An additional benefit is the forgettability of the data. For example, some periods of time may never be shared with anyone, the data is only in the private chain, and some data can be shared with friends or official agencies directly, instead of being broadcast all over the network.

Therefore, payment becomes simple. The buyer is willing to pay, the seller is willing to receive it, and the payment can be completed. With the token-based bookkeeping method, payment can be implemented securely, efficiently, flexibly, and accurately.

This paper proposed an architecture for Web 3.0, and the hashed interactions among nodes can transform bilateral trusts into collective time order. The high transactions per second, low storage costs and flexibility in data management will provide a basis for a new paradigm in individual corporation and social governance, privacy protection and personalization.